\documentclass[fleqn,usenatbib]{mnras}

\usepackage[czech]{babel} % recommended if you write in Czech
\usepackage{soul}

\def\vsini{$V\!\sin i$}

\def\teff{T$_{\rm{eff}}$}
\def\tpole{T$_{\rm{pole}}$}
\def\gpole{$\log~g_{\rm{pole}}$}
\def\geq{$\log~g_{\rm{eq}}$}
\def\teq{T$_{\rm{eq}}$}
\def\rpole{R$_{\rm{p}}$}
\def\req{R$_{\rm{e}}$}
\def\logg{$\log~g$}
\def\degree{$^\circ$\,}

\def\logl{$\log~L/L_{\odot}$}
\def\lage{$\log~{\rm \tau}$}
\def\mmsun{$M/M_{\odot}\,$}
\def\rrsun{$R/R_{\odot}\,$}
\def\rrstar{$R/R_{*}\,$}

\def\omc{$\Omega/\Omega_{\rm{c}}$}
\def\kps{km~s$^{-1}$}
\def\gpc{g~cm$^{-3}$}

\usepackage{newtxtext,newtxmath}
\usepackage[T1]{fontenc}

\DeclareRobustCommand{\VAN}[3]{#2}
\let\VANthebibliography\thebibliography
\def\thebibliography{\DeclareRobustCommand{\VAN}[3]{##3}\VANthebibliography}

\usepackage{graphicx}	% Including figure files
\usepackage{amsmath}	% Advanced maths commands
\usepackage{amssymb}	% Extra maths symbols

\title[The H$\alpha$ line emission of $\beta$ Psc]{The H$\alpha$ line emission of the Be star $\beta$ Psc: the last 40 years}

\author[R. S. Levenhagen et al.]{
Ronaldo S. Levenhagen,$^{1}$\thanks{E-mail: ronaldo.levenhagen@unifesp.br (RSL)}
Marcos P. Diaz$^{2}$,
Eduardo B. Am\^ores$^{3}$,
and Nelson V. Leister $^{2}$
\\
$^{1}$Departamento de F{\'i}sica, Universidade Federal de Sao Paulo, Diadema, Brazil, Rua Prof. Artur Riedel, 275, 09972-270, Diadema, SP, Brazil\\
$^{2}$ Universidade de Sao Paulo, Instituto de Astronomia, Geof{\'i}sica e Ciencias Atmosfericas, Rua do Matao, 1226 Sao Paulo, SP 05508--900, Brazil\\
$^{3}$ Departamento de F{\'i}sica, Universidade Estadual de Feira de Santana (UEFS), Av. Transnordestina, S/N, CEP 44036-900 Feira de Santana, BA, Brazil
}
\date{Accepted 2020 September 08. Received 2020 August 12; in original form 2020 May 16}

\pubyear{2020}

\begin{document}
\label{firstpage}
\pagerange{\pageref{firstpage}--\pageref{lastpage}}
\maketitle

\begin{abstract}
A study on the photosphere and disc of the Be star $\beta$ Psc is presented. We recover almost 40 years 
of high-resolution spectroscopic observations and additional data gathered 
from the BeSS database. We evaluate the photospheric parameters from the SED and fittings of state-of-the-art 
non-LTE model atmospheres to observed helium, carbon, silicon and magnesium line profiles. Our models include the stellar 
geometric deformation as well as the co-latitude dependence of temperature and gravity, aiming to derive 
the effects of rotation on the stellar parameters. We estimate the circumstellar disc parameters from 
the fitting of models assuming different disc properties, namely its radius and gas density profile. 
The disc inclination angle $i$ is constrained from the fittings of He\,{\sc i} 4471 \AA, Mg\,{\sc ii} 4481 \AA, C\,{\sc ii} 4267 \AA\, and Si\,{\sc ii} 4128, 4132 \AA\, lines with gravity darkened models. Our findings, 
based on model fittings, suggest that during the last 40 years the disc radius changed within the 
interval $5.5 \le R_d \le 7.8$ \rrstar, the disc base gas density within $5 \times 10^{-13} \le \rho 
\le 1 \times 10^{-12}$ \gpc, while the radial power-law density index $m$ assumed values between 2.0 
and 2.3. These results are in agreement with recent works dealing with spectroscopic and interferometric measurements of this object.
\end{abstract}


%%%%%%%%%%%%%%%%%%%%%%%%%%%%%%%%%%%%%%%%%%%%%%%%%%

%%%%%%%%%%%%%%%%% BODY OF PAPER %%%%%%%%%%%%%%%%%%

\section{Introduction}

The emission-line B stars, assigned as Be stars, are in general main-sequence stars manifesting Balmer emission lines, sometimes associated to the occurrence of metal emission lines. Many studies today interpret these emission lines as being formed in 
circumstellar thin decretion discs in Keplerian motion around the central star \citep{2013A&ARv..21...69R,2018MNRAS.476.3555R,2020A&A...636A.110D}.

Many models of the circumstellar environment, assuming different geometries, were proposed 
to explain the observed emission profiles and fluxes  
\citep{1969ApJ...156..135M,1972ApJ...171..549H,1986MNRAS.218..761H,1989PASP..101..417C,1993ApJ...409..429B,2000A&A...359.1075H}. 
The line emission varies over a broad time-scale range, which adds to conceal the dynamics of the 
decretion disc. Models of the disc suggest a vast parameter space even during quiescent states \citep{2020ApJ...890...86S}.  

In this work, we study $\beta$ Psc (HR 8773, HD 217891), a bright Be star ($\rm M_{V}=4.5$) that we 
classify as a B6Ve object \citep{2004AJ....127.1176L,2006MNRAS.371..252L}. Its H$\alpha$ emission profile 
is bottle-shaped, and its intensity changes cyclically over the years
\citep{2002ApJ...573..359A,2005A&A...441..235Z,2011AJ....141..150J,2013A&A...550A..79C}.

Recently, \cite{2018ApJ...853..156W} investigated the nature and origin of the rapid rotation of Be stars. 
It becomes more and more accepted that some Be stars acquired their high rotation speeds through the 
mechanism of mass transfer in a close binary system \citep{2008EAS....32..187M}. In this scenario, 
the primary star accretes from a secondary star mass donor which end as a hot, stripped-down 
object. In this case, the system could be eventually identified as a Be+sdO binary. \cite{2018ApJ...853..156W} 
included $\beta$ Psc in their study but they did not detect a possible hot companion through cross-correlation 
analysis of its UV spectra against hot stellar templates. 

Previous studies reveal substantial variations in the H$\alpha$ line flux with respect 
to the local continuum along months/years, as well as long and short-term line
profile variability \citep{1978ApJS...38..205S,1982A&AS...48...93A,1988A&A...189..147H,1996A&AS..116..309H,2009A&A...504..929S}. 
However, the physical properties of the circumstellar disc of this object and their 
correlation with variations in the H$\alpha$  profile remain unknown.

Here we present the results of the analysis of high dispersion spectroscopic (ESO/FEROS) observations 
and extensive synoptic data gathered from the literature. In Section 2, we present detailed information 
on the photometric and spectroscopic data. In Section 3 the physical parameters of the central 
star are estimated assuming that rapid rotation causes the flattening of the stellar poles and 
enlargement at the equator and as a consequence a colatitude-dependent temperature distribution
\citep{1924MNRAS..84..665V,1924MNRAS..84..684V}. In Section 4, we model the H$\alpha$ line 
profiles aiming to estimate the basic parameters of the circumstellar environment for each 
observing epoch, namely the disc radius and gas density distribution. We discuss the results 
obtained in our analysis in Section 5 and the conclusion is shown in Section 6.

\section{Observational Data} \label{sec:obs}

 The photometry data handled in this work come from several sources. We exploited Johnson's 
11-color photometry data by \cite{2002yCat.2237....0D}\footnote{https://cdsarc.unistra.fr/viz-bin/cat/II/237}, 2MASS $JHK_{S}$ photometry 
by \cite{2006AJ....131.1163S}, 13-colour photometry by \cite{1975RMxAA...1..299J} 
and Gaia DR2 photometry \citep{2016A&A...595A...2G,2018A&A...616A...1G}. Also, additional 
photometry data by \cite{2016ApJ...830...51S}, \cite{2012ApJS..199....8G},
\cite{2019A&A...623A..72K}, \cite{2001KFNT...17..409K} and \cite{1978A&AS...34..477M}
were employed.

\begin{figure*}
    \includegraphics[width=13.5cm]{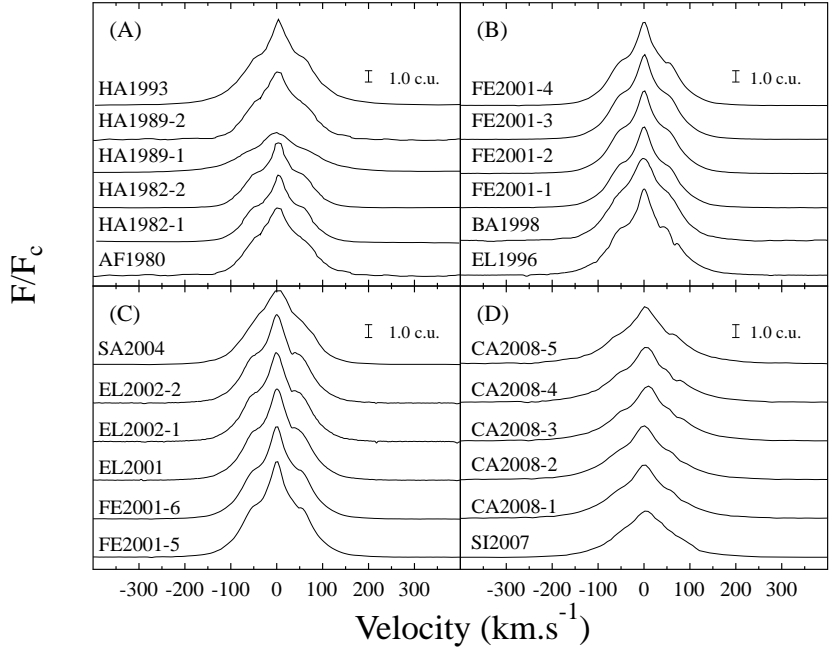}
    \caption{H$\alpha$ line profiles analysed in this work. Each spectrum is assigned to a reference code
shown in Table \ref{tab:1}. The vertical bar represents the continuum unit flux.}
    \label{fig1}
\end{figure*}

We observed the star $\beta$ Psc during six nights, from 2001/Aug/02 to 2001/Oct/09 using the 
FEROS spectrograph \citep{1999Msngr..95....8K} attached to the European Southern Observatory 
(ESO) 1.52 m telescope at La Silla (Chile). Figure \ref{fig1} 
shows the H$\alpha$ line profiles for each night of the run. Given that the H$\alpha$ peak intensity 
and profile remained nearly constant along time-scales of several hours, we analyzed only the 
first spectrum of the observing night as representative of the profile's intensity for that 
night. Table 1 shows the leading information on these data. 

It is worth noticing that even during the whole six nights, the peak intensity remained almost constant. 
The FEROS spectrograph sampled the optical spectrum from 3700 \AA\, to 9000 \AA, with a resolving power 
R$=\frac{\lambda}{\Delta \lambda} \simeq 48,000$  and a typical continuum signal-to-noise ratio of $\sim300$.
The simple optical setup has two fibres with 2''.7 apertures, aimed to record simultaneously the 
incoming stellar flux and sky background. The detector is a back-illuminated CCD with 2948 X 4096 
pixels, with 15 $\mu$m pixel size. 

The data reduction followed standard procedures, with bias and scattered light subtraction, 
Echelle orders extraction, flatfielding, and wavelength calibration.  
We also performed the correction to the local standard of rest and continuum normalization 
with low-order polynomials. We conveyed data reduction with IRAF\footnote{IRAF is 
a data reduction facility that was distributed by NOAO, administrated by the Association 
of Universities for Research in Astronomy (AURA), Inc., under a cooperative agreement 
with the National Science Foundation} package. 

Besides the FEROS  spectroscopic observations, we also analysed the spectroscopic data from the Be Star Spectra 
Database (BeSS)/ELODIE \citep{2004PASP..116..693M,2011AJ....142..149N} containing spectra taken in four epochs at the OHP.
Observing data by several other authors in the literature were also included, as shown in Table 
\ref{tab:1} and in Figure \ref{fig1}. Also, a team of amateur astronomers \footnote{We thank Valerie 
Desnoux, Arnold de Bruin, Andr\'e Favaro, Alun Halsey, Anton Heidemann, Christian Buil, Olivier Thizy, 
Michel Pujol, Carl Sawicki, Erik Bryssinck, Ernst Pollmann, Joan Guarro Fl\'o,
Jean-No\"el Terry, Michel Bonnement, Marco Leonardi, Olivier Garde, Alain Lopez,
Pierre Dubreuil, Robert Buchheim, St\'ephane Ubaud, Thierry Garrel, Thierry Lemoult} 
observed and provided 103 spectra in the BeSS database. We used these
spectra to evaluate H$\alpha$ equivalent widths and peak strengths. The BeSS database is an online catalogue 
that receives spectroscopic data continuously, aiming to include all known Be stars.

\begin{table*}
\caption{Spectroscopic data handled in this work. Each spectra is assigned to a reference code that is used in Figure \ref{fig1}. Amateur's data, also distributed in the BeSS database, are not included in this table.}
\label{tab:1}       % Give a unique label
\centering
\tiny{
\begin{tabular}{l|ccccccc}
\hline\hline
Date  &  No of spectra & MJD & Resolving power & Wavelength coverage (nm) & Instrument & Source & Reference code\\
\hline\hline
 2001/Aug/02 & 15 &  52124 & 48,000 & 370 - 900 & FEROS/1.52m ESO La Silla & this work  & FE2001-1\\
 2001/Aug/05 & 23 &  52127 & 48,000 & 370 - 900 & FEROS/1.52m ESO La Silla & this work  & FE2001-2\\
 2001/Aug/06 & 15 &  52128 & 48,000 & 370 - 900 & FEROS/1.52m ESO La Silla & this work  & FE2001-3\\
 2001/Oct/07 & 12 &  52190 & 48,000 & 370 - 900 & FEROS/1.52m ESO La Silla & this work  & FE2001-4\\
 2001/Oct/08 & 15 &  52191 & 48,000 & 370 - 900 & FEROS/1.52m ESO La Silla & this work  & FE2001-5\\
 2001/Oct/09 & 19 &  52192 & 48,000 & 370 - 900 & FEROS/1.52m ESO La Silla & this work  & FE2001-6\\
\hline
 1996/Sep/02 & 01 &  50328 & 45,000 & 389 - 681 & ELODIE/1.93m OHP & BeSS database  & EL1996\\
 2001/Dec/20 & 01 &  52263 & 45,000 & 389 - 681 & ELODIE/1.93m OHP & BeSS database  & EL2001\\ 
 2002/Nov/30 & 01 &  52608 & 45,000 & 389 - 681 & ELODIE/1.93m OHP & BeSS database  & EL2002-1\\
 2002/Dec/02 & 01 &  52610 & 45,000 & 389 - 681 & ELODIE/1.93m OHP & BeSS database  & EL2002-2\\
\hline
 1975/Dec/05 & 01 &  42751 & 2,200  & 655 - 657 & Photoelectric scanner/72inch Lowell Obs.& {\cite{1978ApJS...38..205S}} & SR1975\\ 
 1976/Nov/05 & 01 &  43087 & 2,200  & 655 - 657 & Photoelectric scanner/72inch Lowell Obs.& {\cite{1978ApJS...38..205S}} & SR1976\\
\hline
 1980/Dec/28 & 01 &  44601 & 10,000 & 644 - 668 & Echelle/1.52m OHP & {\cite{1982A&AS...48...93A}} & AF1980\\  
\hline
 1982/Aug/16  & 01 &  45197  & 100,000 & 655 - 657 & Echelle/1.4m ESO La Silla & {\cite{1988A&A...189..147H}} & HA1982-1\\ 
 1982/Aug/30  & 01 &  45211  & 100,000 & 653 - 659 & Echelle/1.4m ESO La Silla & {\cite{1988A&A...189..147H}} & HA1982-2\\
\hline
 1989/Jan/06  & 01 &  47532  &  50,000 & 655 - 657 & Echelle/1.4m ESO La Silla & {\cite{1996A&AS..116..309H}} & HA1989-1\\
 1989/Sep/26  & 01 &  47795  &  50,000 & 655 - 657 & Echelle/1.4m ESO La Silla & {\cite{1996A&AS..116..309H}} & HA1989-2\\
 1993/Sep/09  & 01 &  49239  &  50,000 & 655 - 657 & Echelle/1.4m ESO La Silla & {\cite{1996A&AS..116..309H}} & HA1993\\
\hline
 1998/Nov/27  & 01 &  51144  &  13,000 & 655 - 658 & FLAGS/Mount Abu IR Observatory & {\cite{2000A&AS..147..229B}} & BA1998\\
\hline 
 2004/Jul/30  & 01 &  53216  &  20,000 & 345 - 862 & HEROS/2.0m Ond\v{r}ejov & {\cite{2006A&A...450..427S}} & SA2004\\ 
\hline
 2007/Dec/18  & 01 &  54452  &  10,000 & 653 - 660 & Fiber-fed Echelle/42inch Lowell Obs. & {\cite{2010ApJS..187..228S}} & SI2007\\
\hline
 2008/Jul/31  & 01 &  54678  &  21,000   &  430 - 680 & FRESCO/91cm INAF-Catania & {\cite{2013A&A...550A..79C}} & CA2008-1\\
 2008/Aug/03  & 01 &  54681  &  21,000   &  430 - 680 & FRESCO/91cm INAF-Catania & {\cite{2013A&A...550A..79C}} & CA2008-2\\
 2008/Sep/04  & 01 &  54713  &  21,000   &  430 - 680 & FRESCO/91cm INAF-Catania & {\cite{2013A&A...550A..79C}} & CA2008-3\\
 2008/Sep/05  & 01 &  54714  &  21,000   &  430 - 680 & FRESCO/91cm INAF-Catania & {\cite{2013A&A...550A..79C}} & CA2008-4\\
 2008/Oct/07  & 01 &  54746  &  21,000   &  430 - 680 & FRESCO/91cm INAF-Catania & {\cite{2013A&A...550A..79C}} & CA2008-5\\
\hline

\end{tabular}
}
\end{table*}

%\begin{figure*}
%\centering
%\includegraphics[width=17cm]{rslevenhagen-01.eps}
%\caption{H$\alpha$ line profiles analysed in this work. Panel (A): spectra by \cite{1982A&AS...48...93A}(a), \cite{1988A&A...189..147H} and \cite{1996A&AS..116..309H} (b). Panel (B): data from FEROS (c). Panel (C): data from \cite{2010ApJS..187..228S} (d) and  \cite{2013A&A...550A..79C} (e). Panel (D): data from BeSS (f), \cite{2000A&AS..147..229B} (g) and \cite{2006A&A...450..427S} (h).}
%\label{fig1}
%\end{figure*}

\section{Photospheric parameters} \label{sec:param}

We estimate the physical parameters of the stellar photosphere (\teff, \logg, \vsini) 
from both photometric measurements and optical spectra.
Vizier2\footnote{https://vizier.u-strasbg.fr/viz-bin/VizieR} \citep{2000A&AS..143...23O}
sourced the currently available photometric data. The fitting of observed broadband 
fluxes with model stellar atmospheres is useful to infer the first-order estimates of
photospheric temperature and stellar radius, provided that reliable information on parallax is available.

Through our preliminary analysis by fitting a black-body, we have selected Johnson's 
11-color photometry data by \cite{2002yCat.2237....0D}, 
2MASS $JHK_{S}$ photometry by \cite{2006AJ....131.1163S}, 13-colour photometry by 
\cite{1975RMxAA...1..299J} and Gaia DR2 photometry \citep{2016A&A...595A...2G,2018A&A...616A...1G}. 
Also, we used additional photometry data by \cite{2016ApJ...830...51S}, \cite{2012ApJS..199....8G},
\cite{2019A&A...623A..72K}, \cite{2001KFNT...17..409K} and \cite{1978A&AS...34..477M}.
Unfortunately, most of the photometric measurements on this star come without uncertainty
estimates.

Preliminary estimate of \vsini\, is derived from the Fourier transforms \citep{1933MNRAS..93..478C, 2005oasp.book.....G}  
of helium line profiles in the spectra, considering the quadratic limb-darkening
coefficients by \cite{1985A&AS...60..471W}.

\begin{figure}
\centering
\includegraphics[width=8.3cm]{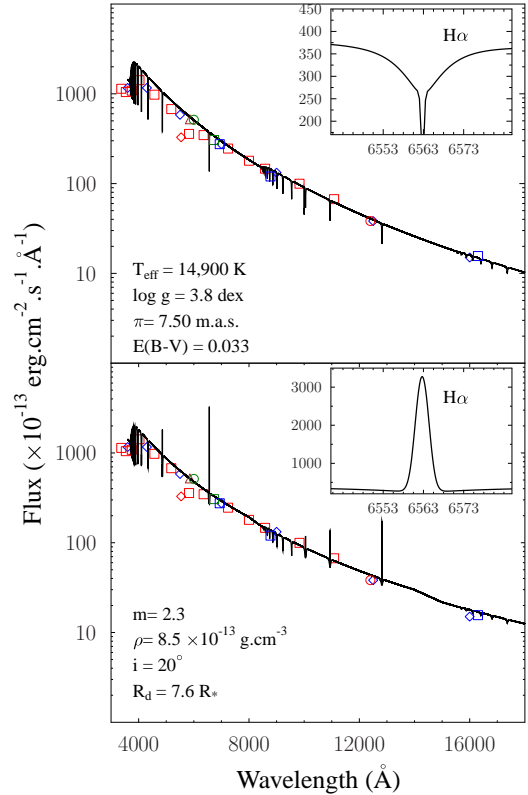}
\caption{Top panel: Fit of synthetic non-LTE TLUSTY+SYNSPEC flux to the SED of 
$\beta$ Psc. The model fluxes were reddened assuming the model A2 published by 
Am\^ores \& Lepine (2007) with E(B-V) $=$ 0.033. The observed SED is corrected by the 
brightening shift induced by geometrical distortion assuming a bilinear interpolation
of Mv $\times$ (B-V) data \protect\citep{2004MNRAS.350..189T}.
%, which makes the star to increase its brightening by 0.38 mag. The stellar parallax was fitted within the limitations imposed by the Gaia DR2 and Hipparcos observational measurements.The best fit solution is achieved for \teff = $14900 \pm 650 $ K, \logg = $3.8 \pm 0.1$ c.g.s. units and parallax $\pi = 7.5 \pm 0.5$ mas. 
Red squares correspond to 13-color 
photometry; Blue diamonds to Johnson's 11-color photometry and 2MASS; Brown triangle is assigned to Gaia DR2 photometry. Green circles stand for data from \protect\cite{2016ApJ...830...51S},
red diamonds are data from \protect\cite{2012ApJS..199....8G}, green squares are photometry by \protect\cite{2019A&A...623A..72K}. Red circles are data observed by \protect\cite{2001KFNT...17..409K}, while blue squares are from \protect\cite{1978A&AS...34..477M}. Top inset shows a zoomed 
H$\alpha$ photospheric profile, with fluxes in linear scale. Bottom panel: Fit of a 
disc flux model plus stellar atmosphere, assuming the same extinction and geometrical 
distortion corrections as before. It was assumed a Keplerian velocity law for the disc (j=0.5), including shear, thermal and disc expanding velocities (see text).  
The fitting solution scales the gas density to $\rho = 8.5\times 10^{-13}$ \gpc, radial 
index $m=2.3$, disc radius $R_d = 7.6 R_{*}$ and inclination angle $i=20$\degree. Bottom inset shows 
the corresponding H$\alpha$ line profile in emission. Flux is in linear scale, in the 
same units as the main plot.}
\label{fig2}
\end{figure}

After this preliminary estimate of the physical state of the photosphere, described 
mainly by its effective temperature \teff, a more detailed parameter set is determined
from the fittings of non-LTE models to the observed SED (Figure \ref{fig2}, top panel)
\citep{2004AJ....127.1176L,2006MNRAS.371..252L,2011NewA...16..307L,2011A&A...533A..75L,2013NewA...18...55L,2013NewA...21...27L}. 
We synthesized non-LTE model spectra using the SYNSPEC v.51 FORTRAN code
\citep{1994A&A...282..151H} covering the spectral range 3000~\AA\, to 18000~\AA\, from
plane-parallel, non-LTE atmosphere models computed with TLUSTY v.205
\citep{1988CoPhC..52..103H}. We used the wrapper SYNPLOT/IDL GUI (Hubeny, priv. comm.) 
to perform the whole synthesis procedure. Solar abundances from
\cite{1998SSRv...85..161G} were considered for both atmosphere structure and spectral synthesis. 

The synthetic fluxes were reddened assuming the A2 model published by \cite{2007AJ....133.1519A}
considering the star's direction, yielding E(B-V) $=$ 0.033. This value 
is an agreement with those 0.031 and 0.027 obtained by \cite{2018MNRAS.478..651G} and
\cite{2003A&A...409..205D} as well as with the obtained one by 
\cite{2017A&A...606A..65C}, e.g., E(B-V) $= 0.026 \pm 0.014$.

Using the \cite{1989ApJ...345..245C}  
extinction curve, we applied the correction factor for all photometric bands from {\it U} to {\it K}. 
The maximum reddening correction along the analyzed SED is 16\% at 3600\AA\, being 9.5\% at 5500\AA. 

When comparing to spherical SED models, the SED was also corrected for the effects of stellar geometrical 
deformation induced by rotation, hereafter geometrical flattening (GF).
The main effect of rotation in a pole-on Be star is to increase the stellar brightening up to
perhaps a half of magnitude \citep{2004MNRAS.350..189T}. In this work we inferred the
correction assuming a bilinear interpolation of Mv $\times$ (B-V) data \citep{2004MNRAS.350..189T}, 
which makes the star to increase its brightening by about 0.38 mag when compared to a spherical photosphere.

The flux fittings resulted in a parallax of $7.5 \pm 0.5$ m.a.s. that provides the
best scaling factor (Figure \ref{fig2}). This value is consistent with the newest Gaia DR2 parallax of $7.71 \pm 0.24$ m.a.s. \citep{2016A&A...595A...2G,2016yCat.1337....0G,2018A&A...616A...9L,2018A&A...616A...1G}.
The older estimate of $6.62 \pm 0.81$ m.a.s. by \citet{Perryman1997} push the fittings to slightly higher 
\teff\,  values.

We performed the SED and spectroscopic fitting procedures with the help of a downhill simplex 
algorithm \citep{Nelder1965} in the vicinity of the starting (\teff,\logg,\vsini)
values.

The goodness of fit was evaluated using the figure-of-merit criterion (FOM) by \cite{1977NucIM.145..389B}:

\begin{equation}
{\rm FOM} = \frac{\sum_i \vert \Phi_{\rm exp} - \Phi_{\rm fit} \vert }{\sum_i \Phi_{\rm fit}}    
\end{equation}

\noindent where $\Phi_{\rm exp}$ corresponds to the experimental data and $\Phi_{\rm fit}$
to the fit. The FOM criterion improves on the uncertainty and fluctuations of the 
$\chi^2$ formula \citep{1977NucIM.145..389B}. A FOM value lower than 2.5\% points out 
a good fit irrespective of variations in line profile shapes and peak sizes
\citep{1977NucIM.145..389B}. 

The evolutionary tracks by \cite{1992A&AS...96..269S}, were interpolated
to estimate the stellar mass \mmsun, the luminosity in log scale \logl and the age in log
scale \lage\, fixing $Z=0.02$.

 The best fit solution is achieved for \teff = $14900 \pm 650 $ K, \logg = $3.8 \pm 0.1$ 
c.g.s. units and parallax $\pi = 7.5 \pm 0.5$ m.a.s., with uncertainties following from the 
probability distributions built using the tabulated data. The complete set of parameters 
from the SED fitting, are shown in the left column of Table \ref{tab:2}.

The high rotation velocities presented by Be stars, which lead the star to a non-spherical oblate 
shape (GF), also affects the observed line profiles in many ways. This rapidly rotating star has 
a polar radius that is significantly smaller 
than the equatorial one. This leads the atmosphere at the polar regions to receive more 
energy per unit area than near the equator 
\citep{1963ApJ...138.1134C,1965ZA.....61..203R}.

Deep into the star, the radiative flux is governed by the temperature gradient, which is related 
to a pressure difference among the inner layers. As the star remains in hydrostatic equilibrium, 
the pressure gradient is ruled by gravity, while the emerging flux is proportional to the local
gravity at photosphere, as stated by Von Zeipel's theorem.

The surface gravity is calculated throughout the star by evaluating  
the local gravitational potential gradient using the Roche approximation with the
inclusion of an additional term for GF \citep{1978trs..book.....T}.
In this approach, we neglect multipole terms in the polynomial expansion that 
arise from non-uniform mass distribution \citep{1966ApJ...146..152C}.
To build-up gravity-darkened (GD+GF) models, we employed the ZPEKTR code described 
in a previous work \citep{2014ApJ...797...29L}.  The computed GD models depart 
from normal non-LTE model spectra. The simulations consider the star as a rigid rotator obeying 
a Roche mass distribution and suppose that the stellar rotation does not influence its core. 
The models are characterized by a set of parameters, such as the surface 
temperature, radius and gravity taken at the poles and equator, the stellar mass, the stellar 
rotation rate \omc, and the aspect angle $i$. 

 Since Be star discs are situated along the equatorial plane, to associate the aspect angle i 
with the disc inclination is a reasonable assumption. For each model, we made a computational 
mesh of thousands of area elements at the stellar surface. Each element has its local parameters $T(\theta)$ and \logg($\theta$), where $\theta$ defines 
the local latitude, in agreement with the Von Zeipel's expression
\citep{1924MNRAS..84..665V,1924MNRAS..84..684V,2006ApJ...643..460L}. The local atmosphere structure 
and radiative transfer is again calculated in non-LTE using TLUSTY and SYNSPEC codes. 
The stellar spectrum is evaluated from the integration of the outgoing specific 
intensities coming from all visible elements in the line-of-sight. The whole set of 
model spectra is used as a base to fit the observed spectra with the Amoeba algorithm \citep{Nelder1965}.

\begin{figure}
\centering
\includegraphics[width=8.3cm]{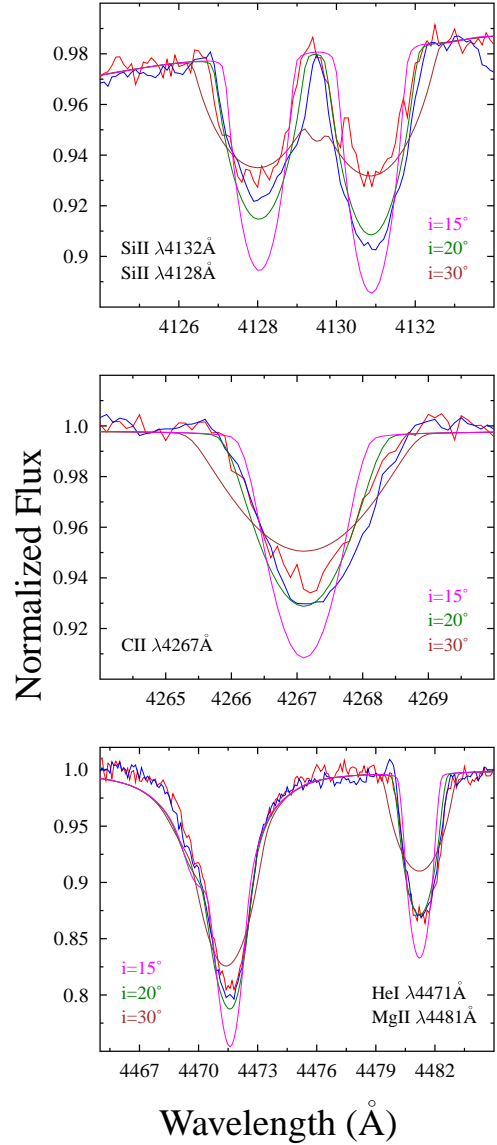}
\caption{Fittings of FEROS (blue) and ELODIE (red) line profiles in three different regions. 
All non-LTE models take into account the GD and GF effects. The best fit models are obtained for $\Omega/\Omega_{\rm c} = 0.79$. Top panel: Si\,{\sc ii} $\lambda\lambda$ 
4128, 4131 \AA\, line profiles in the ELODIE dataset have almost the same central depth, but in
the FEROS dataset the Si\,{\sc ii} $\lambda$ 4131 \AA\, appears strengthened perhaps due to
non-LTE processes in the disc. The model with $i=20$\degree lies in between the others. 
Middle panel: both FEROS and ELODIE C\,{\sc ii} $\lambda$ 4267 \AA\, line have the same 
intensity and are well fitted by a model with $i=20$\degree. Bottom panel: He\,{\sc i} $\lambda$
4471 \AA\, and Mg\,{\sc ii} $\lambda$ 4481 \AA\, line profiles are also well described with 
$i=20$\degree. Higher $i$ values (brown line, $i=30$\degree) do not fit the cores while 
the Mg\,{\sc ii} line becomes wider than observed.}
\label{fig3}
\end{figure}

Figure \ref{fig3} shows the best fits of Si\,{\sc ii} $\lambda\lambda$ 4128, 4131 \AA, 
C\,{\sc ii} $\lambda$ 4267 \AA, He\,{\sc i} $\lambda$ 4471 \AA\, and Mg\,{\sc ii} $\lambda$ 4481 \AA\, 
profiles with gravity-darkened synthetic spectra. The basic parameters derived from these profiles 
are given in the right column of Table \ref{tab:2}. For comparison purposes, we present spectra 
computed considering $i=15$\degree and $i=30$\degree. From these fittings, it is possible to infer 
that the best choice for the aspect angle $i$ is a value near $i=20$\degree. As expected, the bottle-shaped, 
sharp H$\alpha$ emission of $\beta$ Psc is only compatible with a small $i$ 
value. The aspect angle $i=20$\degree inferred from the fittings is perhaps close to its lower 
absolute limit, since values below $i = 20$\degree would lead to a break-up scenario for this star, 
assuming typical parameters of a main sequence B6V star. The H$\alpha$ emission suggests even lower 
angles, although $i=10 - 15$\degree would produce much deeper atmospheric profiles than those observed 
for most lines. These constraints on the inclination should be read with caution since they are limited 
by the assumption of stellar profile formation. It is difficult to precisely establish the value for $i$ 
from the absorption profile fitting since the Mg\,{\sc ii} line appears to be filled in by the disc 
emission. This is possibly the case for the Si\,{\sc ii} and C\,{\sc ii} profiles as well, which 
may be filled in by different amounts.

\begin{table*}
\caption{Stellar parameters from non-LTE models to SED photometry and photospheric line profiles including GF and GD.}
\label{tab:2}       % Give a unique label
\centering

\begin{tabular}{cc|cc}
\hline\hline
  &  SED   &    &  Line profiles \\
\hline\hline
\teff  & $14900 \pm 650$ K   & \tpole    &   $15556 \pm 600$ K      \\
\logg  & $3.8 \pm 0.1$ c.g.s. & \teq     &  $13476 \pm 600$ K      \\
\lage  & $7.92 \pm 0.04$ yr   & \rpole    &  $3.60 \pm 0.21$ \rrsun      \\
\logl  & $3.19 \pm 0.26$      & \req      &  $4.09 \pm 0.21$ \rrsun      \\
\mmsun & $5.0 \pm 0.5$     & \gpole    &  $3.99 \pm 0.14$ c.g.s. \\
       &                   & \geq      &  $3.74 \pm 0.14$ c.g.s. \\     
       &                     & \vsini    & $90 \pm 15$ \kps \\
       &                     & \omc      &  $0.79 \pm 0.13$      \\
       &                     & $i$       &  20\degree $\pm$ 3\degree  \\
\hline\hline
\end{tabular}
\end{table*}

\section{Line emission modelling} \label{sec:emis}

We modelled the emission-line profiles with a modified version of the SHELLSPEC v.39 code
\citep{2004CoSka..34..167B}, following the general outlines given in \cite{2000A&A...359.1075H}, 
where the circumstellar environment is supposed to be described by an axisymmetric disc, using an exponential 
law for the vertical gas density distribution and a radial power-law gradient. Each gas voxel is orbiting the 
star at a radius $R$ with Keplerian velocity given by:

\begin{equation}
V_{kepl} = \left( \frac{GM}{R} \right)^{1/2}    
\end{equation}

\noindent and a projected Doppler velocity shift:

\begin{equation}
V_{D} = V_{kepl} \sin{i} \sin{\theta}    
\end{equation}

\noindent where $i$ stands for the aspect angle ($i=0$ for pole-on view) and $\theta$ the azimuth angle concerning the line-of-sight. 

Since photons originate mostly in an optically thin disc region, the Keplerian shear along the line of sight increases 
the Doppler gradients. We adopt the geometrical prescription for the Doppler gradient from \citet{1986MNRAS.218..761H}:

\begin{equation}
V_{shear} = -\frac{H}{2R}V_{kepl}\sin{i} \tan{i} \sin{\phi} \cos{\phi}    
\end{equation}

\noindent where $H$ is the thickness of the emission layer.

We compute the H$\alpha$ emission profiles by solving the line transfer equation, in LTE conditions, 
along the line-of-sight in the static transfer approximation. An optically thin recombination line 
emission prevails at the circumstellar disc medium density. The total absorption coefficient takes into 
account the H\,{\sc i} bound-free opacity \citep{1978stat.book.....M,2005oasp.book.....G} and 
H\,{\sc i} free-free opacity \citep{1978stat.book.....M}. A pure hydrogen disc is assumed to model 
the Balmer lines. Thomson \citep{1978stat.book.....M} and Rayleigh \citep{1970SAOSR.309.....K} 
processes contribute to the total scattering. The total emission coefficient is considered as the sum of the 
thermal emissivity, computed in LTE, and the scattering emissivity. The monochromatic optical depth 
is given by \citep{2000A&A...359.1075H}:

\begin{equation}
\tau_{\nu} = \frac{\pi e^2}{mc}f \rho(r,z) \frac{\lambda_0}{\sqrt{2\pi}\Delta V \cos{i}} \exp{-\frac{1}{2} \left( \frac{V-V_D}{\Delta V} \right)^2 }  
\end{equation}

\noindent where $f$ stands for the quantum oscillator strength, $\rho(r,z)$ is the gas density distribution, 
and the resulting velocity field $\Delta V$ depends on the thermal broadening, the shear velocity and the 
disc expansion velocity, assuming a radial temperature profile \citep{1994A&A...292..221S}:

\begin{equation}
T_d(R) = T_{\rm{eff}} \left( R/R_*  \right)^{\beta}
\end{equation}

\noindent We compute the emergent flux from the integration, over the solid angle, of the 
outcoming specific intensities from each pixel. At each pixel we evaluate the local physical 
variables (such as the gas density, the thermal profile, the Keplerian and expanding velocities) 
and after that we integrate the optical depths for each frequency point. The solution 
of the radiative transfer equation, reads:

\begin{equation}
I(\tau,\mu) = I(\tau_{\rm layer},\mu)e^{-\frac{(\tau_{\rm layer}-\tau)}{\mu}} + \frac{1}{\mu} \int_{\tau}^{\tau_{\rm layer}} e^{-\frac{(t -\tau)}{\mu}} S(t) dt
\end{equation}

\noindent where $I(\tau_{\rm layer},\mu)$ is the incident intensity of the radiation 
in a layer with optical depth $\tau = \tau_{\rm layer}$. The monochromatic flux is 
$F_{\nu}(x,y) = \int I_{\nu}(x,y) dxdy$ and the circumstellar material is assumed 
to be distributed along a thin disc with inner radius $r_{\rm in}$ and outer 
radius $r_{\rm out}$ with inclination angle $i$. The gas density 
$\rho(r,z)$ follows a radial power-law depending on the distance from the disc inner radius $r$ and
 height from the disc midplane $z$ \citep{2000A&A...359.1075H}:

\begin{equation}
\rho(r,z) = \rho(r_{0}) r^{-m} \exp \left[ -\frac{1}{2} \left( \frac{z}{h(r)} \right)^2 \right]
\end{equation}

\noindent where $m$ is the power-law index for the radial gradient, $\rho(r_{0})$ is the reference gas density and $h(r)$ is the disc scale height, which is a function of the disc
radius, speed of sound $C_s$ and local disc velocity, given by \citep{2000A&A...359.1075H}:

\begin{equation}
h(r)=\frac{C_s}{V_{kepl}} R^{1.5}
\end{equation}

\noindent where $C_s$ follows from the mass and momentum conservation expressions 
and the gas state equation:

\begin{equation}
C_s = \sqrt{\gamma R_g T_d}
\end{equation}

\noindent where $T_d$ stands for the disc temperature, $R_g = 8.31434 \times 10^7 {\rm\, erg.K^{-1}.mole^{-1}}$ 
is the universal gas constant and $\gamma = \frac{5}{3}$ is the adiabatic expansion 
constant for mono-atomic gases. Through the whole fitting process, we fix the disc 
inclination angle $i$ as obtained in the gravity-darkened line profile models, and vary the base 
gas density $\rho$, the power-law index $m$ and the disc radius. Our modelling portrays 
a first-order approximation for the emission line profile, providing constraints on the proposed disc 
parameters. A self-consistent description of the disc temperature and density structure 
is needed to accurately model the profiles.

\section{Discussion} \label{sec:res}

The fitting of photometric fluxes and observed stellar spectra with non-LTE models constrain the stellar 
parameters to  \teff $= 14900 \pm 650$ K, 
\logg $= 3.8 \pm 0.1$ c.g.s. units and \vsini $= 90 \pm 15$ \kps , which is compatible with 
a B6Ve spectral classification (Figure \ref{fig2}). 

This result is in agreement with other studies on this object. 
\cite{1991Ap&SS.183...91T} provide \teff = $15310 \pm 750$ K, while \cite{1978ApJS...38..205S} and 
\cite{1996A&AS..116..309H} suggest \vsini = 100 \kps. 

Higher \teff\, estimates were considered by \cite{2018MNRAS.474.5287A}, who derived \teff\, = $16000 \pm 160$ K,
\logg = $3.00 \pm 0.03$ c.g.s. units, \vsini = $90\pm2$ \kps and assigned a B4V spectral type. Their 
parameters were derived by the fitting of non-LTE model spectra to He\,{\sc i} 4471 \AA\, and Mg\,{\sc ii} 4481 \AA, 
with a best fit reduced $\chi^2$ of 22.6. 

Lower temperature and gravity values were reported by \cite{2005A&A...440..305F}, who 
estimated \teff $= 14359 \pm 295$ K, \logg = $3.672 \pm 0.048$ c.g.s. units and 
\vsini = $95 \pm 5$ \kps as the apparent parameters (i.e. neglecting 
rotation effects) of the star, also derived by the fitting of non-LTE spectra. The different 
set of parameters found for this object in the literature could arise from several reasons, ranging from the accuracy
of the stellar atmospheres models (e.g. number of ions, number of explicit levels, among other 
parameters) to the actual disc activity status. In high active phases, the disc emission
can distort the line profiles and change its core depth, FWHM and wings, even for He\,{\sc i} profiles.

Assuming rigid rotation and taking into account gravity darkening effects, 
the best fitting model has pole temperature \tpole = 15556 K, equator temperature \teq = 13476 K, 
pole gravity \gpole = 3.99 and equator gravity \geq = 3.74 c.g.s. units (Figure \ref{fig3}) with a 
figure-of-merit 0.64\% \citep{1977NucIM.145..389B}. 

This calculated FOM value shows that the observed 
and fitted data are in good agreement. The best-fitting model has an aspect angle $i = 20$\degree and 
leads, for a B6 star, to a equatorial velocity $V_e = 263$ \kps assuming  \vsini = 90 km/s. 

This value is well below the critical limit 
of $V_c = 416 - 418$ km/s for a B6 object \citep{1996MNRAS.280L..31P,2004MNRAS.350..189T}, in agreement 
with \cite{2008ApJ...687..598J}. Lower $i$ values are reported by \cite{2013A&A...550A..79C} 
with $i=10$\degree and \vsini = 75 km/s, implying a critical $V_e = 431$ km/s.  

\begin{figure}
\centering
\includegraphics[width=8.3cm]{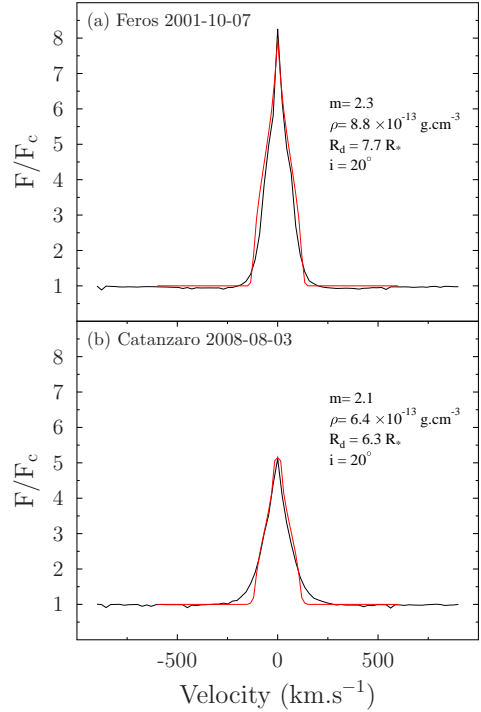}
\caption{Model fitting of H$\alpha$ line profiles with SHELLSPEC. Both models assume $i = 20$\degree and isothermal 
gas with $T_d=0.6$\teff, where \teff = 14,900 K. The fittings are performed over a 3-D parameter space, varying the 
radial exponent $m$, the base gas density $\rho$ and the disc radius $R_{d}$.}
\label{fig4}
\end{figure}

Two representative H$\alpha$ line profiles and their fittings are presented in Figure \ref{fig4} as an example of 
higher and lower emission phases of $\beta$ Psc. The spectra was normalized by the local continuum. In this 
figure, we display the model spectra in red. These spectra observed in 2001 and 2008 were selected for fitting our modified SHELLSPEC models assuming $1 \le m \le 5$, 10\degree $\le i \le$ 50\degree and 
$1.0 \times 10^{-13} \le \rho \le 1.0 \times 10^{-11}$ \gpc. 
The overall line profile fittings are good, except the H$\alpha$ wings that are not always well described by the 
model disc emission (Figure \ref{fig4}, lower panel). We can see significant emission in the H$\alpha$ line 
wings well beyond the maximum expected Keplerian velocity. Such extra emission may be due to scattering of 
emission line photons in a low-density hot circumstellar gas. 

The base gas density in the stellar disc varied within $8.8 - 6.4 \times 10^{-13}$ \gpc and the radial exponent 
$2.1 \le m \le 2.3$, adopting $i=20$\degree. The external disc radius changed from 7.7\rrstar to 6.3\rrstar, 
reflecting variations in mass loss or even viscous processes in the disc.  

\begin{figure}
\centering
\includegraphics[width=8.3cm]{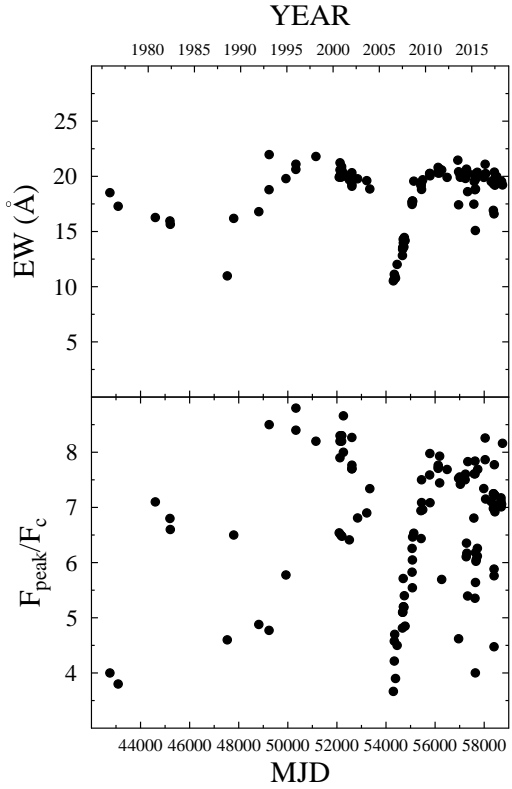}
\caption{Top panel: Long term changes in H$\alpha$
equivalent width from 1975 to 2019. The whole set of spectra is
shown, including all BeSS data. Bottom panel: H$\alpha$ peak strengths relative to local continuum 
from 1975 to 2019.}
\label{fig5}
\end{figure}

Besides our FEROS data, several authors observed the H$\alpha$ line profiles of $\beta$ Psc with 
different instrumentation 
\citep{1978ApJS...38..205S,1982A&AS...48...93A,1988A&A...189..147H,1996A&AS..116..309H,2000A&AS..147..229B,2006A&A...450..427S,2010ApJS..187..228S} 
and recently, many amateur astronomers contributing to the BeSS database. The whole set of observations are not
uniformly sampled in time. Over the last 40 years, one can see many gaps of months or years, which harm
periodicity estimations.  
The H$\alpha$ emission profiles always remain bottle-shaped, but their peak strengths and equivalent
widths changed over the last 40 years (Figure \ref{fig5}). From this figure, it is possible to see only
one interval with a suitable time sample to draw a disc feeding scenario, from the end of 2008 until 
2010, with the shape of an ascending ramp. From that, we could probe a short time scale of a few months 
required to feed the disc. It is worth noticing that the equivalent widths vary between 10 and 20 \AA\, 
with some correlation between equivalent widths and peak strengths, in particular during the transition 
to brighter line emission in 2006. 

\begin{figure}
\centering
\includegraphics[width=8.3cm]{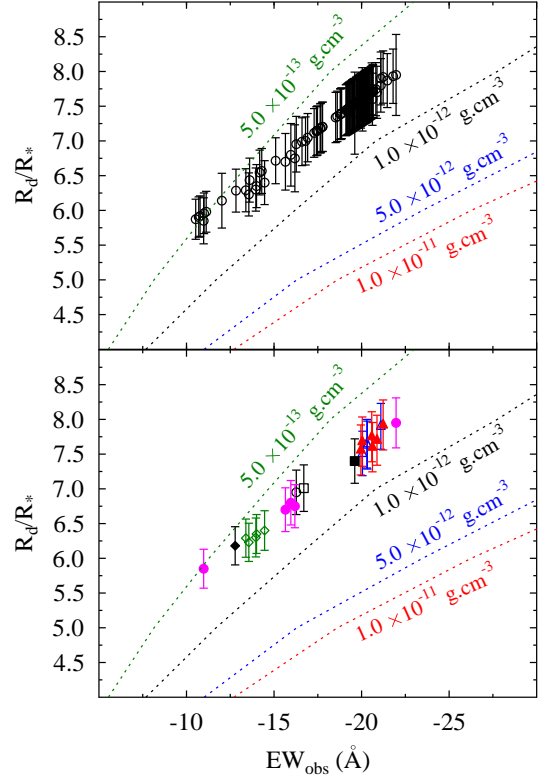}
\caption{Top panel: A comparison of the H$\alpha$ disc radii with observed equivalent widths. All BeSS spectra 
(including the amateur's data) and data from literature are shown.
Low panel: The same comparison, but without amateur's data. Andrillat (black open circle); Banerjee (black open square); Catanzaro (green open diamonds); Elodie (blue open triangles); Feros (red filled triangles); Hanuschik (magenta filled circles); Saad (black filled square); Silaj (black filled diamond).
In all panels, the theoretical models assumed $m=2.0$ and $i=20$\degree, with the following constant gas densities: 
Green dashed line: model with $\rho = 5.0 \times 10^{-13}$ \gpc; Black dashed line: model with constant gas density $1.0 \times 10^{-12}$ \gpc; Blue dashed line: model with $5.0 \times 10^{-12}$ \gpc; Red dashed line: theoretical model with $1.0 \times 10^{-11}$ \gpc. }
\label{fig6}
\end{figure}

These archival data allow the investigation of basic disc parameters over a long time frame. In Figure
\ref{fig6}, we show the range of disc radii as a function of equivalent widths for all data (top panel)
and excluding the lower resolution amateur data (bottom panel). We found a good agreement between those
data sets. Theoretical models with different gas densities $1.0 \times 10^{-12} \le \rho \le 1.0 \times 10^{-11}$ \gpc 
are displayed in both plots. We see that, in both plots, the whole data set fits between 5.5 and 
7.8 \rrstar. Also, the data indicate a gas density region between $5 \times 10^{-13}$ \gpc and 
$1 \times 10^{-12}$ \gpc. 

The gas densities found in this work are in overall agreement with the values found in the literature.
A lower limit of $5.2 \times 10^{-13}$ \gpc\, was obtained by \cite{2013A&A...550A..79C} while in the upper
tail a value of $3 \times 10^{-12}$ \gpc, with $i=20$\degree and $m=2.3$  was found by 
\cite{2008ApJ...687..598J} and \cite{2010ApJS..187..228S} using interferometric observations. 

\begin{figure}
\centering
\includegraphics[width=8.3cm]{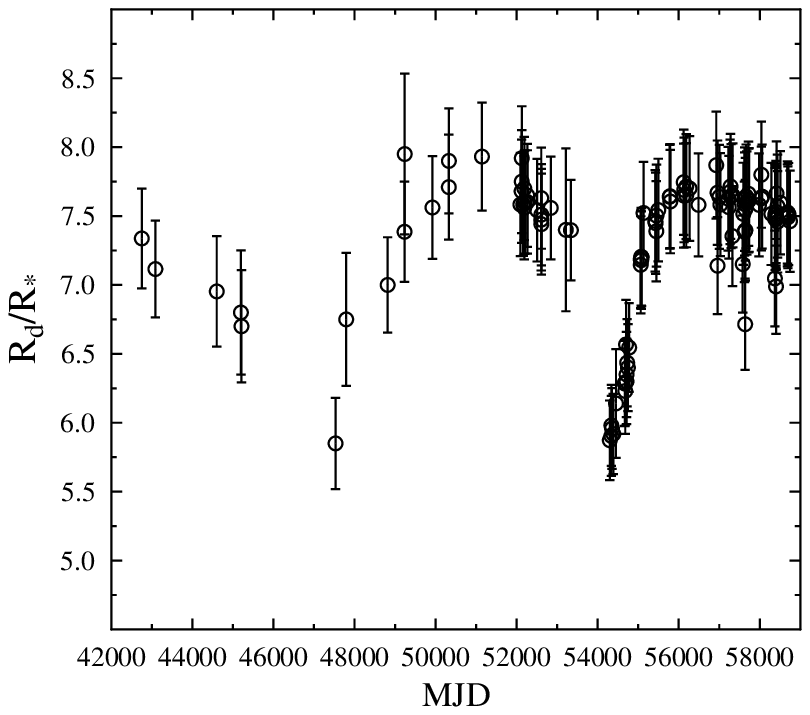}
\caption{Time series of the H$\alpha$ disc radii. All BeSS spectra 
(including the amateur's data) and data from literature are shown. }
\label{fig7}
\end{figure}

The dip seen in Figure \ref{fig5}, both in EW and peak height time series near MJD 54000, and in
Figure \ref{fig7} would suggest the occurrence of stellar brightening changes, 
maybe due to recent ejected material very close to the star, scattering more light to the 
observer. However, by inspecting the Hipparcos light curves we see that the photometry data changed 
only 0.03 mag during the whole mission. As the Hipparcos photometric band is large, it is mainly 
dominated by the continuum. So, the variability we see in peak height and EW is perhaps produced 
mainly by the line and not by the continuum. 
To mimic this stellar brightening effect, we artificially added and subtracted a factor of 0.03 mag
from the continuum and re-normalized the spectra. In this case, small changes of around 1.3\% 
in gas density and 1.2\% in the stellar disc radius are obtained. Supposing, on the contrary, an upper limit case
with around 0.25 mag of variation in the continuum, which is the order of magnitude seen in recent 
photometry by amateurs in the AAVSO database, we would have a fluctuation of about 5\% in the disc radius and 
12\% in the gas density. Nevertheless, these data are heterogeneous and may contain non-intrinsic 
RMS contributions at that level. Roughly one magnitude continuum brightening would be required to 
produce the observed dip in the normalized line peak. However, further indication of relative continuum 
stability regarding the EWs and peak strengths may be found in ASAS-3 \citep{2005AcA....55..275P} 
V-band light curve, where broadband variations below 0.2 mag on this particular time-scale are 
seen during the dip.

\section{Conclusion} \label{sec:con}

We present the analysis of a comprehensive H$\alpha$ line profile data set of the Be star $\beta$ Psc.
Physical parameters of the stellar photosphere are estimated using both photometry data from the literature and 
fittings of line profiles with non-LTE stellar atmospheres models with geometric flattening and gravity darkening. 
We derive a disc inclination of $i = 20^{\circ} \pm 3^{\circ}$. Lower $i$ values showed worse fittings of
C\,{\sc ii}, Si\,{\sc ii}, He\,{\sc i} 4471 \AA\, and Mg\,{\sc ii} 4481 \AA\, line profiles 
and yield equatorial velocities higher than the critical velocity.

Observing the whole H$\alpha$ EW and peak strength data sets, due to the unevenly-spaced time sampling, it was 
not possible to find periodicities in the disc activity. However, the observed line emission variations 
suggest that the disc feeding occurs on timescales of a few months. Simultaneous photometric data suggests 
that most of large EW variations are due to the line emission.

Our fittings of H$\alpha$ with disc models point out that the disc changed its geometrical extent, over the 
last 40 years, ranging between 5.5 to 7.8 \rrstar. The mean base gas density is estimated around 
$7.0 \times 10^{-13}$ \gpc\, and its radial profile is described by a power-law with index between $m=2.0$ and 
$m=2.3$. The model parameters derived from the optical spectra in this work are in reasonable agreement with 
the values proposed in previous works on this object \citep{2008ApJ...687..598J,2013A&A...550A..79C}.

\section*{Acknowledgements}
The authors are grateful to Ivan Hubeny for his valuable advice with the codes SYNSPEC and TLUSTY, and also to Dr. J\'an Budaj for his assistance with the SHELLSPEC code. The authors 
want to acknowledge an anonymous reviewer for valuable comments and suggestions which 
helped to improve this work.
This research handled data from the European Space Agency (ESA) mission
{\it Gaia} (\url{https://www.cosmos.esa.int/gaia}), processed by the {\it Gaia}
Data Processing and Analysis Consortium (DPAC,
\url{https://www.cosmos.esa.int/web/gaia/dpac/consortium}). Funding for the DPAC
has been provided by national institutions, in particular, the institutions
participating in the {\it Gaia} Multilateral Agreement.
This publication made use of data products from
the Two Micron All Sky Survey, which is a joint project of the University
of Massachusetts and the Infrared Processing and Analysis Center/California
Institute of Technology, funded by the National Aeronautics and Space
Administration and the National Science Foundation.
This work has made use of the BeSS database, operated at LESIA, Observatoire de Meudon, 
France: \url{http://basebe.obspm.fr}.
The CNPq (Conselho Nacional de Desenvolvimento 
Cient\'{\i}fico e Tecnol\'ogico) supported this research through grants 307095/2008-8 and 307660/2011-7. 
FAPESP (Funda\c{c}\~ao de Amparo \`a Pesquisa do Estado de S\~ao Paulo) supported this research through 
grant no. 2010/06816-4. The FEROS observations at the European Southern Observatory (ESO) 
were carried out within the Observat\'orio Nacional ON/ESO and ON/IAG agreements, under 
FAPESP project 1998/10138-8.

\section*{Data availability}

The data underlying this article will be shared on reasonable request to the corresponding author.


\newpage

\bibliographystyle{mnras}
\bibliography{reference} % if your bibtex file is called example.bib

\newcommand{\SortNoop}[1]{}
\begin{thebibliography}{}
\makeatletter
\relax
\def\mn@urlcharsother{\let\do\@makeother \do\$\do\&\do\#\do\^\do\_\do\%\do\~}
\def\mn@doi{\begingroup\mn@urlcharsother \@ifnextchar [ {\mn@doi@}
  {\mn@doi@[]}}
\def\mn@doi@[#1]#2{\def\@tempa{#1}\ifx\@tempa\@empty \href
  {http://dx.doi.org/#2} {doi:#2}\else \href {http://dx.doi.org/#2} {#1}\fi
  \endgroup}
\def\mn@eprint#1#2{\mn@eprint@#1:#2::\@nil}
\def\mn@eprint@arXiv#1{\href {http://arxiv.org/abs/#1} {{\tt arXiv:#1}}}
\def\mn@eprint@dblp#1{\href {http://dblp.uni-trier.de/rec/bibtex/#1.xml}
  {dblp:#1}}
\def\mn@eprint@#1:#2:#3:#4\@nil{\def\@tempa {#1}\def\@tempb {#2}\def\@tempc
  {#3}\ifx \@tempc \@empty \let \@tempc \@tempb \let \@tempb \@tempa \fi \ifx
  \@tempb \@empty \def\@tempb {arXiv}\fi \@ifundefined
  {mn@eprint@\@tempb}{\@tempb:\@tempc}{\expandafter \expandafter \csname
  mn@eprint@\@tempb\endcsname \expandafter{\@tempc}}}

\bibitem[\protect\citeauthoryear{{Abt}, {Levato}  \& {Grosso}}{{Abt}
  et~al.}{2002}]{2002ApJ...573..359A}
{Abt} H.~A.,  {Levato} H.,   {Grosso} M.,  2002, \mn@doi [\apj]
  {10.1086/340590}, \href
  {https://ui.adsabs.harvard.edu/abs/2002ApJ...573..359A} {573, 359}

\bibitem[\protect\citeauthoryear{{Am{\^o}res} \& {L{\'e}pine}}{{Am{\^o}res} \&
  {L{\'e}pine}}{2007}]{2007AJ....133.1519A}
{Am{\^o}res} E.~B.,  {L{\'e}pine} J.~R.~D.,  2007, \mn@doi [\aj]
  {10.1086/511418}, \href
  {https://ui.adsabs.harvard.edu/abs/2007AJ....133.1519A} {133, 1519}

\bibitem[\protect\citeauthoryear{{Andrillat} \& {Fehrenbach}}{{Andrillat} \&
  {Fehrenbach}}{1982}]{1982A&AS...48...93A}
{Andrillat} Y.,  {Fehrenbach} C.,  1982, \aaps, \href
  {https://ui.adsabs.harvard.edu/abs/1982A&AS...48...93A} {48, 93}

\bibitem[\protect\citeauthoryear{{Arcos}, {Kanaan}, {Ch{\'a}vez}, {Vanzi},
  {Araya}  \& {Cur{\'e}}}{{Arcos} et~al.}{2018}]{2018MNRAS.474.5287A}
{Arcos} C.,  {Kanaan} S.,  {Ch{\'a}vez} J.,  {Vanzi} L.,  {Araya} I.,
  {Cur{\'e}} M.,  2018, \mn@doi [\mnras] {10.1093/mnras/stx3075}, \href
  {https://ui.adsabs.harvard.edu/abs/2018MNRAS.474.5287A} {474, 5287}

\bibitem[\protect\citeauthoryear{{Balian} \& {Eddy}}{{Balian} \&
  {Eddy}}{1977}]{1977NucIM.145..389B}
{Balian} H.~G.,  {Eddy} N.~W.,  1977, \mn@doi [Nuclear Instruments and Methods]
  {10.1016/0029-554X(77)90437-2}, \href
  {https://ui.adsabs.harvard.edu/abs/1977NucIM.145..389B} {145, 389}

\bibitem[\protect\citeauthoryear{{Banerjee}, {Rawat}  \&
  {Janardhan}}{{Banerjee} et~al.}{2000}]{2000A&AS..147..229B}
{Banerjee} D.~P.~K.,  {Rawat} S.~D.,   {Janardhan} P.,  2000, \mn@doi [\aaps]
  {10.1051/aas:2000299}, \href
  {https://ui.adsabs.harvard.edu/abs/2000A&AS..147..229B} {147, 229}

\bibitem[\protect\citeauthoryear{{Bjorkman} \& {Cassinelli}}{{Bjorkman} \&
  {Cassinelli}}{1993}]{1993ApJ...409..429B}
{Bjorkman} J.~E.,  {Cassinelli} J.~P.,  1993, \mn@doi [\apj] {10.1086/172676},
  \href {https://ui.adsabs.harvard.edu/abs/1993ApJ...409..429B} {409, 429}

\bibitem[\protect\citeauthoryear{{Budaj} \& {Richards}}{{Budaj} \&
  {Richards}}{2004}]{2004CoSka..34..167B}
{Budaj} J.,  {Richards} M.~T.,  2004, Contributions of the Astronomical
  Observatory Skalnate Pleso, \href
  {https://ui.adsabs.harvard.edu/abs/2004CoSka..34..167B} {34, 167}

\bibitem[\protect\citeauthoryear{{Capitanio}, {Lallement}, {Vergely},
  {Elyajouri}  \& {Monreal-Ibero}}{{Capitanio}
  et~al.}{2017}]{2017A&A...606A..65C}
{Capitanio} L.,  {Lallement} R.,  {Vergely} J.~L.,  {Elyajouri} M.,
  {Monreal-Ibero} A.,  2017, \mn@doi [\aap] {10.1051/0004-6361/201730831},
  \href {https://ui.adsabs.harvard.edu/abs/2017A&A...606A..65C} {606, A65}

\bibitem[\protect\citeauthoryear{{Cardelli}, {Clayton}  \& {Mathis}}{{Cardelli}
  et~al.}{1989}]{1989ApJ...345..245C}
{Cardelli} J.~A.,  {Clayton} G.~C.,   {Mathis} J.~S.,  1989, \mn@doi [\apj]
  {10.1086/167900}, \href
  {https://ui.adsabs.harvard.edu/abs/1989ApJ...345..245C} {345, 245}

\bibitem[\protect\citeauthoryear{{Carroll}}{{Carroll}}{1933}]{1933MNRAS..93..478C}
{Carroll} J.~A.,  1933, \mn@doi [\mnras] {10.1093/mnras/93.7.478}, \href
  {https://ui.adsabs.harvard.edu/abs/1933MNRAS..93..478C} {93, 478}

\bibitem[\protect\citeauthoryear{{Catanzaro}}{{Catanzaro}}{2013}]{2013A&A...550A..79C}
{Catanzaro} G.,  2013, \mn@doi [\aap] {10.1051/0004-6361/201220357}, \href
  {https://ui.adsabs.harvard.edu/abs/2013A&A...550A..79C} {550, A79}

\bibitem[\protect\citeauthoryear{{Cidale} \& {Ringuelet}}{{Cidale} \&
  {Ringuelet}}{1989}]{1989PASP..101..417C}
{Cidale} L.~S.,  {Ringuelet} A.~E.,  1989, \mn@doi [\pasp] {10.1086/132443},
  \href {https://ui.adsabs.harvard.edu/abs/1989PASP..101..417C} {101, 417}

\bibitem[\protect\citeauthoryear{{Collins}}{{Collins}}{1963}]{1963ApJ...138.1134C}
{Collins} George~W. I.,  1963, \mn@doi [\apj] {10.1086/147712}, \href
  {https://ui.adsabs.harvard.edu/abs/1963ApJ...138.1134C} {138, 1134}

\bibitem[\protect\citeauthoryear{{Collins} \& {Harrington}}{{Collins} \&
  {Harrington}}{1966}]{1966ApJ...146..152C}
{Collins} George~W. I.,  {Harrington} J.~P.,  1966, \mn@doi [\apj]
  {10.1086/148866}, \href
  {https://ui.adsabs.harvard.edu/abs/1966ApJ...146..152C} {146, 152}

\bibitem[\protect\citeauthoryear{{\SortNoop{De}}de~Almeida
  et~al.,}{{\SortNoop{De}}de~Almeida et~al.}{2020}]{2020A&A...636A.110D}
{\SortNoop{De}}de~Almeida E.~S.~G.~d.,  et~al., 2020, \mn@doi [\aap]
  {10.1051/0004-6361/01936039}, \href
  {https://ui.adsabs.harvard.edu/abs/2020A&A...636A.110D} {636, A110}

\bibitem[\protect\citeauthoryear{{Drimmel}, {Cabrera-Lavers}  \&
  {L{\'o}pez-Corredoira}}{{Drimmel} et~al.}{2003}]{2003A&A...409..205D}
{Drimmel} R.,  {Cabrera-Lavers} A.,   {L{\'o}pez-Corredoira} M.,  2003, \mn@doi
  [\aap] {10.1051/0004-6361:20031070}, \href
  {https://ui.adsabs.harvard.edu/abs/2003A&A...409..205D} {409, 205}

\bibitem[\protect\citeauthoryear{{Ducati}}{{Ducati}}{2002}]{2002yCat.2237....0D}
{Ducati} J.~R.,  2002, VizieR Online Data Catalog, \href
  {https://ui.adsabs.harvard.edu/abs/2002yCat.2237....0D} {}

\bibitem[\protect\citeauthoryear{{Fr{\'e}mat}, {Zorec}, {Hubert}  \&
  {Floquet}}{{Fr{\'e}mat} et~al.}{2005}]{2005A&A...440..305F}
{Fr{\'e}mat} Y.,  {Zorec} J.,  {Hubert} A.~M.,   {Floquet} M.,  2005, \mn@doi
  [\aap] {10.1051/0004-6361:20042229}, \href
  {https://ui.adsabs.harvard.edu/abs/2005A&A...440..305F} {440, 305}

\bibitem[\protect\citeauthoryear{{Gaia Collaboration}}{{Gaia
  Collaboration}}{2016}]{2016yCat.1337....0G}
{Gaia Collaboration} 2016, VizieR Online Data Catalog, \href
  {https://ui.adsabs.harvard.edu/abs/2016yCat.1337....0G} {p. I/337}

\bibitem[\protect\citeauthoryear{{Gaia Collaboration} et~al.,}{{Gaia
  Collaboration} et~al.}{2016}]{2016A&A...595A...2G}
{Gaia Collaboration} et~al., 2016, \mn@doi [\aap]
  {10.1051/0004-6361/201629512}, \href
  {https://ui.adsabs.harvard.edu/abs/2016A&A...595A...2G} {595, A2}

\bibitem[\protect\citeauthoryear{{Gaia Collaboration} et~al.,}{{Gaia
  Collaboration} et~al.}{2018}]{2018A&A...616A...1G}
{Gaia Collaboration} et~al., 2018, \mn@doi [\aap]
  {10.1051/0004-6361/201833051}, \href
  {https://ui.adsabs.harvard.edu/abs/2018A&A...616A...1G} {616, A1}

\bibitem[\protect\citeauthoryear{{Gray}}{{Gray}}{2005}]{2005oasp.book.....G}
{Gray} D.~F.,  2005, {The Observation and Analysis of Stellar Photospheres}.
Cambridge University Press

\bibitem[\protect\citeauthoryear{{Green} et~al.,}{{Green}
  et~al.}{2018}]{2018MNRAS.478..651G}
{Green} G.~M.,  et~al., 2018, \mn@doi [\mnras] {10.1093/mnras/sty1008}, \href
  {https://ui.adsabs.harvard.edu/abs/2018MNRAS.478..651G} {478, 651}

\bibitem[\protect\citeauthoryear{{Grevesse} \& {Sauval}}{{Grevesse} \&
  {Sauval}}{1998}]{1998SSRv...85..161G}
{Grevesse} N.,  {Sauval} A.~J.,  1998, \mn@doi [\ssr]
  {10.1023/A:1005161325181}, \href
  {https://ui.adsabs.harvard.edu/abs/1998SSRv...85..161G} {85, 161}

\bibitem[\protect\citeauthoryear{{Gudennavar}, {Bubbly}, {Preethi}  \&
  {Murthy}}{{Gudennavar} et~al.}{2012}]{2012ApJS..199....8G}
{Gudennavar} S.~B.,  {Bubbly} S.~G.,  {Preethi} K.,   {Murthy} J.,  2012,
  \mn@doi [\apjs] {10.1088/0067-0049/199/1/8}, \href
  {https://ui.adsabs.harvard.edu/abs/2012ApJS..199....8G} {199, 8}

\bibitem[\protect\citeauthoryear{{Hanuschik}, {Kozok}  \& {Kaiser}}{{Hanuschik}
  et~al.}{1988}]{1988A&A...189..147H}
{Hanuschik} R.~W.,  {Kozok} J.~R.,   {Kaiser} D.,  1988, \aap, \href
  {https://ui.adsabs.harvard.edu/abs/1988A&A...189..147H} {189, 147}

\bibitem[\protect\citeauthoryear{{Hanuschik}, {Hummel}, {Sutorius}, {Dietle}
  \& {Thimm}}{{Hanuschik} et~al.}{1996}]{1996A&AS..116..309H}
{Hanuschik} R.~W.,  {Hummel} W.,  {Sutorius} E.,  {Dietle} O.,   {Thimm} G.,
  1996, \aaps, \href {https://ui.adsabs.harvard.edu/abs/1996A&AS..116..309H}
  {116, 309}

\bibitem[\protect\citeauthoryear{{Horne} \& {Marsh}}{{Horne} \&
  {Marsh}}{1986}]{1986MNRAS.218..761H}
{Horne} K.,  {Marsh} T.~R.,  1986, \mn@doi [\mnras] {10.1093/mnras/218.4.761},
  \href {https://ui.adsabs.harvard.edu/abs/1986MNRAS.218..761H} {218, 761}

\bibitem[\protect\citeauthoryear{{Huang}}{{Huang}}{1972}]{1972ApJ...171..549H}
{Huang} S.-S.,  1972, \mn@doi [\apj] {10.1086/151309}, \href
  {https://ui.adsabs.harvard.edu/abs/1972ApJ...171..549H} {171, 549}

\bibitem[\protect\citeauthoryear{{Hubeny}}{{Hubeny}}{1988}]{1988CoPhC..52..103H}
{Hubeny} I.,  1988, \mn@doi [Computer Physics Communications]
  {10.1016/0010-4655(88)90177-4}, \href
  {https://ui.adsabs.harvard.edu/abs/1988CoPhC..52..103H} {52, 103}

\bibitem[\protect\citeauthoryear{{Hubeny}, {Hummer}  \& {Lanz}}{{Hubeny}
  et~al.}{1994}]{1994A&A...282..151H}
{Hubeny} I.,  {Hummer} D.~G.,   {Lanz} T.,  1994, \aap, \href
  {https://ui.adsabs.harvard.edu/abs/1994A&A...282..151H} {282, 151}

\bibitem[\protect\citeauthoryear{{Hummel} \& {Vrancken}}{{Hummel} \&
  {Vrancken}}{2000}]{2000A&A...359.1075H}
{Hummel} W.,  {Vrancken} M.,  2000, \aap, \href
  {https://ui.adsabs.harvard.edu/abs/2000A&A...359.1075H} {359, 1075}

\bibitem[\protect\citeauthoryear{{Johnson} \& {Mitchell}}{{Johnson} \&
  {Mitchell}}{1975}]{1975RMxAA...1..299J}
{Johnson} H.~L.,  {Mitchell} R.~I.,  1975, \rmxaa, \href
  {https://ui.adsabs.harvard.edu/abs/1975RMxAA...1..299J} {1, 299}

\bibitem[\protect\citeauthoryear{{Jones}, {Tycner}, {Sigut}, {Benson}  \&
  {Hutter}}{{Jones} et~al.}{2008}]{2008ApJ...687..598J}
{Jones} C.~E.,  {Tycner} C.,  {Sigut} T.~A.~A.,  {Benson} J.~A.,   {Hutter}
  D.~J.,  2008, \mn@doi [\apj] {10.1086/591726}, \href
  {https://ui.adsabs.harvard.edu/abs/2008ApJ...687..598J} {687, 598}

\bibitem[\protect\citeauthoryear{{Jones}, {Tycner}  \& {Smith}}{{Jones}
  et~al.}{2011}]{2011AJ....141..150J}
{Jones} C.~E.,  {Tycner} C.,   {Smith} A.~D.,  2011, \mn@doi [\aj]
  {10.1088/0004-6256/141/5/150}, \href
  {https://ui.adsabs.harvard.edu/abs/2011AJ....141..150J} {141, 150}

\bibitem[\protect\citeauthoryear{{Kaufer}, {Stahl}, {Tubbesing},
  {N{\o}rregaard}, {Avila}, {Francois}, {Pasquini}  \& {Pizzella}}{{Kaufer}
  et~al.}{1999}]{1999Msngr..95....8K}
{Kaufer} A.,  {Stahl} O.,  {Tubbesing} S.,  {N{\o}rregaard} P.,  {Avila} G.,
  {Francois} P.,  {Pasquini} L.,   {Pizzella} A.,  1999, The Messenger, \href
  {https://ui.adsabs.harvard.edu/abs/1999Msngr..95....8K} {95, 8}

\bibitem[\protect\citeauthoryear{{Kervella}, {Arenou}, {Mignard}  \&
  {Th{\'e}venin}}{{Kervella} et~al.}{2019}]{2019A&A...623A..72K}
{Kervella} P.,  {Arenou} F.,  {Mignard} F.,   {Th{\'e}venin} F.,  2019, \mn@doi
  [\aap] {10.1051/0004-6361/201834371}, \href
  {https://ui.adsabs.harvard.edu/abs/2019A&A...623A..72K} {623, A72}

\bibitem[\protect\citeauthoryear{{Kharchenko}}{{Kharchenko}}{2001}]{2001KFNT...17..409K}
{Kharchenko} N.~V.,  2001, Kinematika i Fizika Nebesnykh Tel, \href
  {https://ui.adsabs.harvard.edu/abs/2001KFNT...17..409K} {17, 409}

\bibitem[\protect\citeauthoryear{{Kurucz}}{{Kurucz}}{1970}]{1970SAOSR.309.....K}
{Kurucz} R.~L.,  1970, SAO Special Report, \href
  {https://ui.adsabs.harvard.edu/abs/1970SAOSR.309.....K} {309}

\bibitem[\protect\citeauthoryear{{Levenhagen}}{{Levenhagen}}{2014}]{2014ApJ...797...29L}
{Levenhagen} R.~S.,  2014, \mn@doi [\apj] {10.1088/0004-637X/797/1/29}, \href
  {https://ui.adsabs.harvard.edu/abs/2014ApJ...797...29L} {797, 29}

\bibitem[\protect\citeauthoryear{{Levenhagen} \& {K{\"u}nzel}}{{Levenhagen} \&
  {K{\"u}nzel}}{2011}]{2011NewA...16..307L}
{Levenhagen} R.~S.,  {K{\"u}nzel} R.,  2011, \mn@doi [\na]
  {10.1016/j.newast.2010.09.003}, \href
  {https://ui.adsabs.harvard.edu/abs/2011NewA...16..307L} {16, 307}

\bibitem[\protect\citeauthoryear{{Levenhagen} \& {Leister}}{{Levenhagen} \&
  {Leister}}{2004}]{2004AJ....127.1176L}
{Levenhagen} R.~S.,  {Leister} N.~V.,  2004, \mn@doi [\aj] {10.1086/381063},
  \href {https://ui.adsabs.harvard.edu/abs/2004AJ....127.1176L} {127, 1176}

\bibitem[\protect\citeauthoryear{{Levenhagen} \& {Leister}}{{Levenhagen} \&
  {Leister}}{2006}]{2006MNRAS.371..252L}
{Levenhagen} R.~S.,  {Leister} N.~V.,  2006, \mn@doi [\mnras]
  {10.1111/j.1365-2966.2006.10655.x}, \href
  {https://ui.adsabs.harvard.edu/abs/2006MNRAS.371..252L} {371, 252}

\bibitem[\protect\citeauthoryear{{Levenhagen}, {Leister}  \&
  {K{\"u}nzel}}{{Levenhagen} et~al.}{2011}]{2011A&A...533A..75L}
{Levenhagen} R.~S.,  {Leister} N.~V.,   {K{\"u}nzel} R.,  2011, \mn@doi [\aap]
  {10.1051/0004-6361/201116590}, \href
  {https://ui.adsabs.harvard.edu/abs/2011A&A...533A..75L} {533, A75}

\bibitem[\protect\citeauthoryear{{Levenhagen}, {K{\"u}nzel}  \&
  {Leister}}{{Levenhagen} et~al.}{2013a}]{2013NewA...18...55L}
{Levenhagen} R.~S.,  {K{\"u}nzel} R.,   {Leister} N.~V.,  2013a, \mn@doi [\na]
  {10.1016/j.newast.2012.06.003}, \href
  {https://ui.adsabs.harvard.edu/abs/2013NewA...18...55L} {18, 55}

\bibitem[\protect\citeauthoryear{{Levenhagen}, {K{\"u}nzel}  \&
  {Leister}}{{Levenhagen} et~al.}{2013b}]{2013NewA...21...27L}
{Levenhagen} R.~S.,  {K{\"u}nzel} R.,   {Leister} N.~V.,  2013b, \mn@doi [\na]
  {10.1016/j.newast.2012.10.004}, \href
  {https://ui.adsabs.harvard.edu/abs/2013NewA...21...27L} {21, 27}

\bibitem[\protect\citeauthoryear{{Lovekin}, {Deupree}  \& {Short}}{{Lovekin}
  et~al.}{2006}]{2006ApJ...643..460L}
{Lovekin} C.~C.,  {Deupree} R.~G.,   {Short} C.~I.,  2006, \mn@doi [\apj]
  {10.1086/501492}, \href
  {https://ui.adsabs.harvard.edu/abs/2006ApJ...643..460L} {643, 460}

\bibitem[\protect\citeauthoryear{{Luri} et~al.,}{{Luri}
  et~al.}{2018}]{2018A&A...616A...9L}
{Luri} X.,  et~al., 2018, \mn@doi [\aap] {10.1051/0004-6361/201832964}, \href
  {https://ui.adsabs.harvard.edu/abs/2018A&A...616A...9L} {616, A9}

\bibitem[\protect\citeauthoryear{{Marlborough}}{{Marlborough}}{1969}]{1969ApJ...156..135M}
{Marlborough} J.~M.,  1969, \mn@doi [\apj] {10.1086/149954}, \href
  {https://ui.adsabs.harvard.edu/abs/1969ApJ...156..135M} {156, 135}

\bibitem[\protect\citeauthoryear{{Meynet}}{{Meynet}}{2008}]{2008EAS....32..187M}
{Meynet} G.,  2008, in {Charbonnel} C.,  {Zahn} J.~P.,  eds,  EAS Publications
  Series Vol. 32, EAS Publications Series. pp 187--232 (\mn@eprint {arXiv}
  {0708.3185}), \mn@doi{10.1051/eas:0832006}

\bibitem[\protect\citeauthoryear{{Mihalas}}{{Mihalas}}{1978}]{1978stat.book.....M}
{Mihalas} D.,  1978, {Stellar atmospheres}.
{W H Freeman \& Co.}

\bibitem[\protect\citeauthoryear{{Morel} \& {Magnenat}}{{Morel} \&
  {Magnenat}}{1978}]{1978A&AS...34..477M}
{Morel} M.,  {Magnenat} P.,  1978, \aaps, \href
  {https://ui.adsabs.harvard.edu/abs/1978A&AS...34..477M} {34, 477}

\bibitem[\protect\citeauthoryear{{Moultaka}, {Ilovaisky}, {Prugniel}  \&
  {Soubiran}}{{Moultaka} et~al.}{2004}]{2004PASP..116..693M}
{Moultaka} J.,  {Ilovaisky} S.~A.,  {Prugniel} P.,   {Soubiran} C.,  2004,
  \mn@doi [\pasp] {10.1086/422177}, \href
  {https://ui.adsabs.harvard.edu/abs/2004PASP..116..693M} {116, 693}

\bibitem[\protect\citeauthoryear{{Neiner}, {de Batz}, {Cochard}, {Floquet},
  {Mekkas}  \& {Desnoux}}{{Neiner} et~al.}{2011}]{2011AJ....142..149N}
{Neiner} C.,  {de Batz} B.,  {Cochard} F.,  {Floquet} M.,  {Mekkas} A.,
  {Desnoux} V.,  2011, \mn@doi [\aj] {10.1088/0004-6256/142/5/149}, \href
  {https://ui.adsabs.harvard.edu/abs/2011AJ....142..149N} {142, 149}

\bibitem[\protect\citeauthoryear{{Nelder} \& {Mead}}{{Nelder} \&
  {Mead}}{1965}]{Nelder1965}
{Nelder} J.~A.,  {Mead} R.,  1965, \mn@doi [Computer Journal]
  {https://doi.org/10.1093/comjnl/7.4.308}, 7, 308

\bibitem[\protect\citeauthoryear{{Ochsenbein}, {Bauer}  \&
  {Marcout}}{{Ochsenbein} et~al.}{2000}]{2000A&AS..143...23O}
{Ochsenbein} F.,  {Bauer} P.,   {Marcout} J.,  2000, \mn@doi [\aaps]
  {10.1051/aas:2000169}, \href
  {https://ui.adsabs.harvard.edu/abs/2000A&AS..143...23O} {143, 23}

\bibitem[\protect\citeauthoryear{{Perryman} et~al.,}{{Perryman}
  et~al.}{1997}]{Perryman1997}
{Perryman} M. A.~C.,  et~al., 1997, \aap, 323, 49

\bibitem[\protect\citeauthoryear{{Pojmanski}, {Pilecki}  \&
  {Szczygiel}}{{Pojmanski} et~al.}{2005}]{2005AcA....55..275P}
{Pojmanski} G.,  {Pilecki} B.,   {Szczygiel} D.,  2005, \actaa, \href
  {https://ui.adsabs.harvard.edu/abs/2005AcA....55..275P} {55, 275}

\bibitem[\protect\citeauthoryear{{Porter}}{{Porter}}{1996}]{1996MNRAS.280L..31P}
{Porter} J.~M.,  1996, \mn@doi [\mnras] {10.1093/mnras/280.3.L31}, \href
  {https://ui.adsabs.harvard.edu/abs/1996MNRAS.280L..31P} {280, L31}

\bibitem[\protect\citeauthoryear{{R{\'\i}mulo} et~al.,}{{R{\'\i}mulo}
  et~al.}{2018}]{2018MNRAS.476.3555R}
{R{\'\i}mulo} L.~R.,  et~al., 2018, \mn@doi [\mnras] {10.1093/mnras/sty431},
  \href {https://ui.adsabs.harvard.edu/abs/2018MNRAS.476.3555R} {476, 3555}

\bibitem[\protect\citeauthoryear{{Rivinius}, {Carciofi}  \&
  {Martayan}}{{Rivinius} et~al.}{2013}]{2013A&ARv..21...69R}
{Rivinius} T.,  {Carciofi} A.~C.,   {Martayan} C.,  2013, \mn@doi [\aapr]
  {10.1007/s00159-013-0069-0}, \href
  {https://ui.adsabs.harvard.edu/abs/2013A&ARv..21...69R} {21, 69}

\bibitem[\protect\citeauthoryear{{Roxburgh}, {Griffith}  \& {Sweet}}{{Roxburgh}
  et~al.}{1965}]{1965ZA.....61..203R}
{Roxburgh} I.~W.,  {Griffith} J.~S.,   {Sweet} P.~A.,  1965, \zap, \href
  {https://ui.adsabs.harvard.edu/abs/1965ZA.....61..203R} {61, 203}

\bibitem[\protect\citeauthoryear{{Saad} et~al.,}{{Saad}
  et~al.}{2006}]{2006A&A...450..427S}
{Saad} S.~M.,  et~al., 2006, \mn@doi [\aap] {10.1051/0004-6361:20041877}, \href
  {https://ui.adsabs.harvard.edu/abs/2006A&A...450..427S} {450, 427}

\bibitem[\protect\citeauthoryear{{Schaller}, {Schaerer}, {Meynet}  \&
  {Maeder}}{{Schaller} et~al.}{1992}]{1992A&AS...96..269S}
{Schaller} G.,  {Schaerer} D.,  {Meynet} G.,   {Maeder} A.,  1992, \aaps, \href
  {https://ui.adsabs.harvard.edu/abs/1992A&AS...96..269S} {96, 269}

\bibitem[\protect\citeauthoryear{{Silaj}, {Jones}, {Tycner}, {Sigut}  \&
  {Smith}}{{Silaj} et~al.}{2010}]{2010ApJS..187..228S}
{Silaj} J.,  {Jones} C.~E.,  {Tycner} C.,  {Sigut} T.~A.~A.,   {Smith} A.~D.,
  2010, \mn@doi [\apjs] {10.1088/0067-0049/187/1/228}, \href
  {https://ui.adsabs.harvard.edu/abs/2010ApJS..187..228S} {187, 228}

\bibitem[\protect\citeauthoryear{{Skrutskie} et~al.,}{{Skrutskie}
  et~al.}{2006}]{2006AJ....131.1163S}
{Skrutskie} M.~F.,  et~al., 2006, \mn@doi [\aj] {10.1086/498708}, \href
  {https://ui.adsabs.harvard.edu/abs/2006AJ....131.1163S} {131, 1163}

\bibitem[\protect\citeauthoryear{{Slettebak} \& {Reynolds}}{{Slettebak} \&
  {Reynolds}}{1978}]{1978ApJS...38..205S}
{Slettebak} A.,  {Reynolds} R.~C.,  1978, \mn@doi [\apjs] {10.1086/190554},
  \href {https://ui.adsabs.harvard.edu/abs/1978ApJS...38..205S} {38, 205}

\bibitem[\protect\citeauthoryear{{Stee} \& {de Araujo}}{{Stee} \& {de
  Araujo}}{1994}]{1994A&A...292..221S}
{Stee} P.,  {de Araujo} F.~X.,  1994, \aap, \href
  {https://ui.adsabs.harvard.edu/abs/1994A&A...292..221S} {292, 221}

\bibitem[\protect\citeauthoryear{{{S}tefl} et~al.,}{{{S}tefl}
  et~al.}{2009}]{2009A&A...504..929S}
{{S}tefl} S.,  et~al., 2009, \mn@doi [\aap] {10.1051/0004-6361/200811573},
  \href {https://ui.adsabs.harvard.edu/abs/2009A&A...504..929S} {504, 929}

\bibitem[\protect\citeauthoryear{{Straatman} et~al.,}{{Straatman}
  et~al.}{2016}]{2016ApJ...830...51S}
{Straatman} C. M.~S.,  et~al., 2016, \mn@doi [\apj]
  {10.3847/0004-637X/830/1/51}, \href
  {https://ui.adsabs.harvard.edu/abs/2016ApJ...830...51S} {830, 51}

\bibitem[\protect\citeauthoryear{{Suffak}, {Jones}, {Tycner}, {Henry},
  {Carciofi}, {Mota}  \& {Rubio}}{{Suffak} et~al.}{2020}]{2020ApJ...890...86S}
{Suffak} M.~W.,  {Jones} C.~E.,  {Tycner} C.,  {Henry} G.~W.,  {Carciofi}
  A.~C.,  {Mota} B.~C.,   {Rubio} A.~C.,  2020, \mn@doi [\apj]
  {10.3847/1538-4357/ab68dc}, \href
  {https://ui.adsabs.harvard.edu/abs/2020ApJ...890...86S} {890, 86}

\bibitem[\protect\citeauthoryear{{Tassoul}}{{Tassoul}}{1978}]{1978trs..book.....T}
{Tassoul} J.-L.,  1978, {Theory of rotating stars}.
Princeton University Press

\bibitem[\protect\citeauthoryear{{Theodossiou} \& {Danezis}}{{Theodossiou} \&
  {Danezis}}{1991}]{1991Ap&SS.183...91T}
{Theodossiou} E.,  {Danezis} E.,  1991, \mn@doi [\apss] {10.1007/BF00643019},
  \href {https://ui.adsabs.harvard.edu/abs/1991Ap&SS.183...91T} {183, 91}

\bibitem[\protect\citeauthoryear{{Townsend}, {Owocki}  \& {Howarth}}{{Townsend}
  et~al.}{2004}]{2004MNRAS.350..189T}
{Townsend} R.~H.~D.,  {Owocki} S.~P.,   {Howarth} I.~D.,  2004, \mn@doi
  [\mnras] {10.1111/j.1365-2966.2004.07627.x}, \href
  {https://ui.adsabs.harvard.edu/abs/2004MNRAS.350..189T} {350, 189}

\bibitem[\protect\citeauthoryear{{\SortNoop{Von}}von~Zeipel}{{\SortNoop{Von}}von~Zeipel}{1924a}]{1924MNRAS..84..665V}
{\SortNoop{Von}}von~Zeipel H.,  1924a, \mn@doi [\mnras]
  {10.1093/mnras/84.9.665}, \href
  {https://ui.adsabs.harvard.edu/abs/1924MNRAS..84..665V} {84, 665}

\bibitem[\protect\citeauthoryear{{\SortNoop{Von}}von~Zeipel}{{\SortNoop{Von}}von~Zeipel}{1924b}]{1924MNRAS..84..684V}
{\SortNoop{Von}}von~Zeipel H.,  1924b, \mn@doi [\mnras]
  {10.1093/mnras/84.9.684}, \href
  {https://ui.adsabs.harvard.edu/abs/1924MNRAS..84..684V} {84, 684}

\bibitem[\protect\citeauthoryear{{Wade} \& {Rucinski}}{{Wade} \&
  {Rucinski}}{1985}]{1985A&AS...60..471W}
{Wade} R.~A.,  {Rucinski} S.~M.,  1985, \aaps, \href
  {https://ui.adsabs.harvard.edu/abs/1985A&AS...60..471W} {60, 471}

\bibitem[\protect\citeauthoryear{{Wang}, {Gies}  \& {Peters}}{{Wang}
  et~al.}{2018}]{2018ApJ...853..156W}
{Wang} L.,  {Gies} D.~R.,   {Peters} G.~J.,  2018, \mn@doi [\apj]
  {10.3847/1538-4357/aaa4b8}, \href
  {https://ui.adsabs.harvard.edu/abs/2018ApJ...853..156W} {853, 156}

\bibitem[\protect\citeauthoryear{{Zorec}, {Fr{\'e}mat}  \& {Cidale}}{{Zorec}
  et~al.}{2005}]{2005A&A...441..235Z}
{Zorec} J.,  {Fr{\'e}mat} Y.,   {Cidale} L.,  2005, \mn@doi [\aap]
  {10.1051/0004-6361:20053051}, \href
  {https://ui.adsabs.harvard.edu/abs/2005A&A...441..235Z} {441, 235}

\makeatother
\end{thebibliography}

%\begin{thebibliography}{99}
%\end{thebibliography}

%%%%%%%%%%%%%%%%%%%%%%%%%%%%%%%%%%%%%%%%%%%%%%%%%%

%%%%%%%%%%%%%%%%% APPENDICES %%%%%%%%%%%%%%%%%%%%%

%\appendix



%%%%%%%%%%%%%%%%%%%%%%%%%%%%%%%%%%%%%%%%%%%%%%%%%%

% Don't change these lines
\bsp	% typesetting comment
\label{lastpage}
\end{document}